\documentclass[10pt,false,letter]{article}

\usepackage{fullpage}
\usepackage{setspace}
\usepackage{parskip}
\usepackage{titlesec}
\usepackage{xcolor}
\usepackage{lineno}
\usepackage[export]{adjustbox}

\PassOptionsToPackage{hyphens}{url}
\usepackage[colorlinks = true,
            linkcolor = blue,
            urlcolor  = blue,
            citecolor = blue,
            anchorcolor = blue]{hyperref}
\usepackage{etoolbox}
\makeatletter
\patchcmd\@combinedblfloats{\box\@outputbox}{\unvbox\@outputbox}{}{%
  \errmessage{\noexpand\@combinedblfloats could not be patched}%
}%
\makeatother

\renewenvironment{abstract}
  {{\bfseries\noindent{\large\abstractname}\par\nobreak}}
  {}

\titlespacing{\section}{0pt}{*3}{*1}
\titlespacing{\subsection}{0pt}{*2}{*0.5}
\titlespacing{\subsubsection}{0pt}{*1.5}{0pt}

\usepackage{authblk}

\usepackage{graphicx}
\usepackage[space]{grffile}
\usepackage{latexsym}
\usepackage{textcomp}
\usepackage{longtable}
\usepackage{tabulary}
\usepackage{booktabs,array,multirow}
\usepackage{amsfonts,amsmath,amssymb}
\providecommand\citet{\cite}
\providecommand\citep{\cite}

\newif\iflatexml\latexmlfalse

\DeclareGraphicsExtensions{.pdf,.PDF,.png,.PNG,.jpg,.JPG,.jpeg,.JPEG}

\usepackage[utf8]{inputenc}
\usepackage[english]{babel}

\def\rd{{\rm d}}

\newcommand{\diff}[1]{\frac{\mathrm{d}}{\mathrm{d}#1}}

\newcommand{\dif}[1]{\,\mathrm{d} #1}

\newcommand{\deriv}[2]{\frac{\mathrm{d} #1}{\mathrm{d} #2}}

\begin{document}


\title{Time-Dependent Saddle Node Bifurcation: Breaking Time and the Point of No Return in a Non-Autonomous Model of Critical Transitions}
\author{Jeremiah H. Li$^1$, Felix X.-F. Ye$^2$,
Hong Qian$^1$, and Sui Huang$^3$\\
$^1$Department of Applied Mathematics\\ 
University of Washington\\
Seattle, WA 98195-3925, USA\\
$^2$Department of Applied Mathematics \& Statistics\\
Johns Hopkins University\\
Baltimore, MD 21218, USA\\
$^3$Institute for Systems Biology\\
Seattle, WA 98109, USA
}

\maketitle

\begin{abstract}
There is a growing awareness that catastrophic phenomena in biology and medicine can be mathematically represented in terms of saddle-node bifurcations. In particular, the term ``tipping'', or critical transition has in recent years entered the discourse of the general public in relation to ecology, medicine, and public health. The saddle-node bifurcation and its associated theory of catastrophe as put forth by Thom and Zeeman has seen applications in a wide range of fields including molecular biophysics, mesoscopic physics, and climate science. In this paper, we investigate a simple model of a non-autonomous system with a time-dependent parameter $p(\tau)$ and its corresponding ``dynamic'' (time-dependent) saddle-node bifurcation by the modern theory of non-autonomous dynamical systems. We show that the actual point of no return for a system undergoing tipping can be significantly delayed in comparison to the {\em breaking time} $\hat{\tau}$ at which the corresponding autonomous system with a time-independent parameter $p_{a}= p(\hat{\tau})$ undergoes a bifurcation. A dimensionless parameter $\alpha=\lambda p_0^3V^{-2}$ is introduced, in which $\lambda$ is the curvature of the autonomous saddle-node bifurcation according to parameter $p(\tau)$, which has an initial value of $p_{0}$ and a constant rate of change $V$. We find that the breaking time $\hat{\tau}$ is always less than the actual point of no return $\tau^*$ after which the critical transition is irreversible; specifically, the relation $\tau^*-\hat{\tau}\simeq 2.338(\lambda V)^{-\frac{1}{3}}$ is analytically obtained. For a system with a small $\lambda V$, there exists a significant window of opportunity $(\hat{\tau},\tau^*)$ during which rapid reversal of the environment can save the system from catastrophe.
\end{abstract}

\section{Introduction}

Recently, there has been an intense interest in using a dynamical systems approach to the characterization of a class of phenomenon known as ``critical transitions'', or ``tipping points'' \cite{fritjofcapra}, in clinical medicine \cite{chenluonan,tpincancer-1,tpincancer-2}, ecology \cite{jgore1,jgore2}, environmental and climate science \cite{tpines,scheffer-09}, and sociology \cite{schelling}.  In all these problems, complex systems experience a qualitatively large and abrupt change very rapidly. Moreover, the control parameters in the system may not vary slowly.   Specifically in connection to applications, 
Chen {\it et al.}  based on high-throughput data studied the early-warning 
signals for the onset of two complex diseases, chronic
hepatitis B liver cancer and lung injury induced by carbonyl chloride inhalation \cite{chenluonan}. Dai {\it et al.} investigated in a laboratory
population collapse of the budding yeast {\it Saccharomyces cerevisiae}
due to the Allee effect by continuous dilution  \cite{jgore1} and 
reducing the sucrose concentration in growth medium \cite{jgore2}.
Clinicians were also starting to apply related ideas in cancer therapies \cite{tpincancer-1,tpincancer-2}. Held, Suarez and Fraedrich 
\cite{held-suarez,fraedrich} described the evolution of the global mean surface temperature of an ocean-covered spherical planet subjected to radiative heating which took into account ice-albedo and greenhouse feedback. 
 One of the particular emphases in these studies is the ``model free'', robust character of a critical
transition that is independent of specific
 systems. See \cite{scheffer-09} for a comprehensive review on even more diverse phenomena ranging from epileptic seizure to crisis in financial markets.

The generic nature of this class of phenomena suggests certain
universal mathematical principles \cite{scheffer-09,kuehn-11}.
 For autonomous dynamics, the concepts of attractors, bifurcations and related timescales from the mathematical theory of nonlinear dynamics \cite{strogatz,perko} have been essential in understanding biological systems in mechanistic terms \cite{murray}, with particular successes in fields such as cellular biochemistry \cite{qian-arbp}, developmental biology \cite{huang-ptrsb}, evolutionary genetics \cite{Felix2013} and cancer medicine \cite{huang-pbmb}.  The most commonly encountered one-dimensional bifurcations are of the saddle-node (fold), transcritical, and pitchfork types \cite{strogatz,perko}. There is, however, an essential difference in the robustness, e.g., a tolerance for imperfect bifurcation,
between the latter two types and the first \cite{nicolis-95}. 
In mathematical terms, the latter two types of bifurcations are not structurally stable. In biological terms, saddle node bifurcation is robust nonlinear phenomenon \cite{kuznetsov-13}\footnote{Note that robust pitchfork bifurcations can exist if defined in particular ways, such as the type defined via the ``hidden'' second-order phase transition associated with any cusp catastrophe \cite{aqtw}.}. It is for precisely this reason that saddle-node bifurcations, and the associated cusp catastrophe, play important roles in biology.  Indeed, they have afforded fundamental insights and conceptual understanding of a wide range of biological phenomena, from voltage-dependent channel kinetics and forced macromolecular bond rupture \cite{qian-ps}, to \emph{E.\ coli} phenotype switching \cite{ge-qian-xie-prl} and  ecological insect outbreak \cite{kot}. In addition, saddle-node bifurcations are characterized by the existence of ``bottlenecks", in the side which the saddle node solution do not exist. For systems behave like $\frac{\rd x}{\rd t}=-x^2-p^2$ with small $p$, the trajectory of the dynamics is slowed down or “bottlenecked” by a ``saddle node ghost" \cite{strogatz,kuehn-09}. The time needed to blow up is well dominated by the bottleneck and can be analytically found $t_b= \frac{1}{p}\left[\arctan\left(\frac{x_{0}}{p}\right)+\frac{\pi}{2}\right]$ where $x_0$ is the initial point.

Mathematically, systems undergoing critical transitions can be represented as dynamical systems with saddle-node bifurcations, where the tipping points are the corresponding fold catastrophes. However, it is critical to understand that in real-world systems, the bifurcation parameter of a dynamical system can itself be \emph{time-dependent}; thus, tipping points are fold catastrophes \emph{induced} by parameters changing in time. In other words, tipping points are induced by changes in external conditions, so systems with tipping points should be described by non-autonomous dynamical systems. To be more mathematically precise, these systems can in general be described by a non-autonomous dynamical system
\begin{align}
  \label{eq:nonautonomous-definition}
  \deriv{\vec{x}}{\tau} = \vec{f}(\vec{x},p(\tau)),\ p(\tau)\in \mathbb{R}
\end{align}
such that:
\begin{itemize}
\item If $p(\tau)$ is a constant, the vector field $\vec{f}(\vec{x},p)$ has (for some value $\hat{p}$) an attractor for $p<\hat{p}$ which 
disappears for $p>\hat{p}$.

\item If $p(\tau)$ depends on $\tau$, we have $p(\tau)$ such that: (\emph{i}) at $\tau=\hat{\tau}$, $p(\hat{\tau})=\hat{p}$, (\emph{ii}) $p(\tau)<\hat{p}$ when $\tau<\hat{\tau}$, and (\emph{iii}) $p(\tau)>\hat{p}$ when $\tau>\hat{\tau}$.
\end{itemize}

When the parameters change slowly enough, the system can be approximated as its autonomous counterpart (that is, where $p$ is the time-independent bifurcation parameter), or as a slow-fast system \cite{kuehn-11}. However, cases where such approximations are not valid require treatments of the fully non-autonomous system, of which there are fewer \cite{ashwin-15}. The long-term behavior of non-autonomous system is characterized by the well developed modern mathematical notion of ``pullback attractors". The mathematical setup of pullback attractors is not quite intuitive. Many theoretical approaches focus on the existence and uniqueness of it \cite{caraballo2017}. In present paper, we provided many explicit examples of these pullback attractors analytically and numerically. 
In particular, we emphasized a non-autonomous treatment of a simple system with a tipping point. We do this by introducing an explicitly time-dependent mathematical model for systems with a generic tipping point inspired  by the biology experiments \cite{jgore1,jgore2} and the climate science \cite{held-suarez,fraedrich} . It is called  the dynamical saddle-node bifurcation. Previous results focused on the case of slow-varying parameter and didn't connect to modern theories of non-autonomous dynamical systems \cite{Erneux1989,Haberman1979,Rachel2015}. 
 We also introduce the concepts of a system's ``breaking time''  and ``point of no return'' with the help of the pullback attractor.
The bottleneck feature in autonomous systems is reflected in  a relevant property of the dynamical saddle-node bifurcation: there is a time window between the breaking time and the point of no return. 
Most notably, we discuss how a non-autonomous treatment of tipping point phenomena reveals possibilities for saving systems which autonomous treatments would deem as having already undergone an irreversible transition. To wit: consider the system described in Eq.\ \ref{eq:nonautonomous-definition}: in the autonomous limit where the constant parameter $p$ is varied at an infinitely slow speed, the system undergoes a critical transition precisely at the moment when $p$ crosses $\hat{p}$. However, we show that a non-autonomous treatment reveals that if the parameter begins at $p<\hat{p}$ and increases at a constant rate $V$ to lead to a disappearance of a stable steady state at $\tau=\hat{\tau}$ (which we call the ``breaking time''), the system can be rescued if a rapid reversal of the environment is effected, \emph{even for} $\tau>\hat{\tau}$. However, there is only a finite window of opportunity: a ``point of no return'' $\tau^{*}>\hat{\tau}$ exists , at which a catastrophe occurs. After the point of no return,  the system cannot be saved. This window of opportunity $(\hat{\tau},\tau^{*})$ therefore has potential for great practical significance in managing tipping point phenomena. We obtain the result for the relationship between these two quantities based on the analytic solution for the pullback attractor of our simple model (Eq.\ \ref{eq:model-untrans} below):  $\tau^{*}-\hat{\tau}\simeq 2.338(\lambda V)^{-1/3}$, where
$\lambda$ is the curvature of the autonomous saddle-node bifurcation according to parameter $p$.

The paper is structured as follows. In Sec.\ \ref{sec:non-auton-syst}, we give a brief introduction to non-autonomous systems and the notion of pullback attractors, followed by several simple examples to illustrate the dynamics of non-autonomous systems with: (\emph{i}) a ``moving fixed point'', (\emph{ii}) a moving fixed point with a changing eigenvalue, and (\emph{iii}) a more complex situation with a ``hidden catastrophe''. This section is included purely for pedagogical purposes. Motivated readers are also referred to a more mathematical discussion \cite{kloeden-rasmussen,kloeden-pot}. Sec.\ \ref{sec:dynamic-saddle-node} introduces the main model for the dynamic saddle-node bifurcation (Eq.\ \ref{eq:model-untrans}). We explore its behavior using numerical computation. The notions of $\hat{t}$ and $t^{*}$ are defined, and a relation between them is observed from numerical computations.  We then present the analytical solution of the model in terms of the first Airy function, from which we obtain an exact relation between $\hat{t}$ and $t^{*}$.  The paper concludes with Sec.\ \ref{sec:discuss} which discusses our model and the concepts of breaking time and the point of no return.

\section{Simple non-autonomous dynamics 
and pullback attractors}
\label{sec:non-auton-syst}

\subsection{Pullback attractors and tipping point phenomena}
\label{sec:pullb-attr-tipp}
An autonomous dynamical system described by the system of ordinary differential equations 
\begin{align}
  \label{eq:autonomous-definition}
  \diff{t}\vec{x}(t) = \vec{f}\big(\vec{x}(t),p\big) \qquad \vec{x},\vec{f}(\vec{x})\in\mathbb{R}^{n},t\in\mathbb{R}
\end{align}
has an $n$-dimensional vector field $\vec{f}(\vec{x},p)$ that is independent of time. In the theory of dynamical systems of this form, the concept of attractors is among the most important---they characterize the long-time behavior of a system \cite{strogatz,perko,nicolis-95}. In the context of tipping point phenomena modeled by autonomous systems with saddle node bifurcations, a system always tracks the stable equilibrium point as the parameter is slowly increased, until the the fold catastrophe, at which point the system undergoes a ``critical transition'' (that is, it ``tips'') into another attractor.  

For non-autonomous dynamical systems, however, which are described by a system of ordinary differential equations 
\begin{align}
  \label{eq:nonautonomous-definition2}
  \diff{t}\vec{x}(t) &=  \vec{f}\big(\vec{x}(t),p(t)\big)
        =\vec{g}(\vec{x}(t),t) \qquad \vec{x},\vec{g}(\vec{x},t)\in\mathbb{R}^{n},t\in\mathbb{R}
\end{align}
the vector field $\vec{g}$ is \emph{itself} an explicit function of time---in physical terms, the changing vector field reflects the changing environment of the system. Because of this, a system starting in a particular initial state $\vec{x}^{*}\in\mathbb{R}^{n}$ where $\vec{g}(\vec{x}^{*},t)=0$ when $t=t_{0}$ will in general not remain at $\vec{x}^{*}$ for all future times $t=t_{1}>t_{0}$---contrast this to equilibrium points in autonomous systems, where trajectories starting there remain there for all future time.

Thus, the idea of attractors from autonomous theory is insufficient to describe non-autonomous systems (and therefore general tipping phenomena)---the concepts must be generalized---which is where the concept of a ``pullback attractor'' comes in \cite{kloeden-rasmussen}. Note the fact that the solutions to non-autonomous systems explicitly depend \emph{not only} on the elapsed time $t-t_{0}$ but on \emph{both} $t$ and $t_{0}$. The dependence on two variables thus means there are two types of attractors for non-autonomous systems---in fact, it means that attractors in non-autonomous systems involve a time-dependence, unlike attractors in autonomous systems \cite{kloeden-pot}. We refer to the asymptotic behavior $t\rightarrow +\infty$ as the forward behavior and $t_0 \rightarrow -\infty$ as the pullback behavior. In an autonomous system, both forward and backward are the same, but they are different for non-autonomous systems. Also, the mathematical theory of non-autonomous dynamics has established that attraction of the ``pullback'' type is of a more fundamental character. Because it possesses nice properties, such as if it exists then is unique and it can be seen as a generalization of the global attractor from autonomous theory. But the difficulty arises with the definitions of forward attractors which the existence and uniqueness are not unclear \cite{Robinson2012attractors}.

Qualitatively, a pullback attractor of a non-autonomous system can be thought of as the attractor for a large collection of trajectories which began infinitely long ago; quantitatively, it corresponds with the trajectory of the system with an initial condition $(t_{0},\vec{x}_{0})$ in the case when $t_{0}\to -\infty$. To be more mathematically precise: let $\vec{\varphi}(t\mid \vec{x}_{0},t_{0})$ be the solution to Eq.\ \ref{eq:nonautonomous-definition2} with an initial condition $\vec{x}(t_{0})=\vec{x}_{0}$. Then the pullback attractor of the system is given by $\vec{\varphi}_{pb}(t)$:\footnote{Actually, an even more mathematically precise statement is that the limit of 
$\vec{\varphi}(t|B,t_0)$ as $t_{0}\to -\infty$ for any subset $B\subset\mathbb{R}^n$,
is itself a set. In this paper we only consider the unique global pullback attracting solution, then
we just have the simple Eq.\ \ref{eq:pullback-attractor-definition}.
}
\begin{align}
  \label{eq:pullback-attractor-definition}
  \vec{\varphi}_{pb}(t) \equiv \lim_{t_{0}\to -\infty}\vec{\varphi}(t\mid\vec{x}_{0},t_{0}) 
\end{align}
We consider the case that the pullback attractor is independent of $x_{0}$---in essence, the memory of starting point of the system is lost after enough time elapsed. 
 For systems which start sometime in the finite past, they quickly converge to $\vec{\varphi}_{pb}(t)$, as illustrated in Fig. \ref{fig:1} and Fig. \ref{fig:2}.

What follows are several simple examples. 

\subsection{A moving fixed point}
\label{sec:appr-moving-fixed}

The first example is of a non-autonomous dynamical system with a single fixed point whose location changes smoothly in time; that is, $\deriv{x}{t}=-\lambda(x-\tilde{x}(t))$ where $\tilde{x}(t)$ is the location of the fixed point and $\lambda>0$. The exact solution to this non-autonomous ODE is 
\begin{align}
  \label{eq:moving-fixed-point1}
  \varphi(t\mid x_{0},t_{0}) &= x_{0}e^{-\lambda(t-t_{0})}+\lambda\int_{t_{0}}^{t}e^{-\lambda(t-\tau)}\tilde{x}(\tau)\dif{\tau}
  \\
\label{eq:moving-fixed-point2}
  &= \left(x_{0}-\tilde{x}(t_{0})\right)e^{-\lambda(t-t_{0})}+\tilde{x}(t)-\int_{0}^{t-t_{0}}e^{-\lambda s}\left(\deriv{\tilde{x}(t-s)}{t}\right)\dif{s}. 
\end{align}
We obtain the pullback attractor by taking $t_{0}\to -\infty$:
\begin{align}
  \label{eq:moving-fixed-point-pullback}
  \varphi_{pb}(t) = \lambda\int_{-\infty}^{t}e^{-\lambda(t-\tau)}\tilde{x}(\tau)\dif{\tau} = \lambda\int_{0}^{\infty}e^{-\lambda s}\tilde{x}(t-s)\dif{s}.
\end{align}
It is instructive to note that $\varphi_{pb}(t)$ differs from the function one obtains if one takes $t\to\infty$ in Eq.\ \ref{eq:moving-fixed-point1}: 
\begin{align}
  \label{eq:moving-forward}
  \varphi_{f}(t_{0}) &=\lim_{t\to\infty}\lambda e^{-\lambda t}\int_{t_{0}}^{t}e^{\lambda\tau}\tilde{x}(\tau)\dif{\tau} = \lim_{\tau\to\infty}\lambda\int_{0}^{\tau}e^{-\lambda s}\tilde{x}(\tau+t_{0}-s)\dif{s}.
\end{align}
$\varphi_{f}(t_{0})$ describes the forward behavior \cite{kloeden-rasmussen,Robinson2012attractors}. If $\tilde{x}(t)\equiv\tilde{x}_{0}$ is independent of $t$, then both Eqs. \ref{eq:moving-fixed-point-pullback} and \ref{eq:moving-forward} are equal to the same constant $\tilde{x}_{0}$. If $\tilde{x}(t)$ has limits $\tilde{x}_{\pm\infty}$ when $t$ tends to $\pm\infty$, then $\varphi_{f}(t_{0})=\tilde{x}_{+\infty}$, but $\varphi_{pb}(t)$ is a smooth function of $t$ that connects $\tilde{x}_{-\infty}$ to $\tilde{x}_{+\infty}$ \cite{kloeden-pot}. In general, however, it is not certain that the limit in Eq.\ \ref{eq:moving-forward} always exists, although the pullback attractor \emph{itself} always exists.

\begin{figure}
\begin{tabular}{ccc}
 \includegraphics[width=0.33\linewidth]{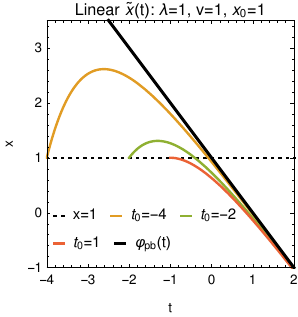} &
 \includegraphics[width=0.33\linewidth]{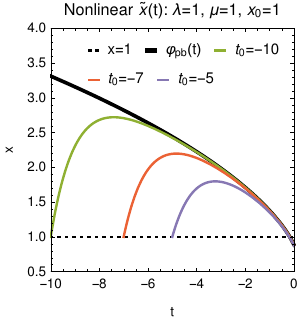} &
\includegraphics[ width=0.33\linewidth]{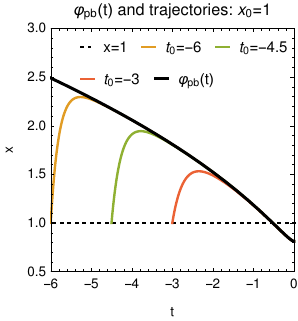}\\
\bf{\Large (A)} &\bf{\Large (B)} &\bf{\Large (C) }
\end{tabular}
\caption{Trajectories and pullback attractors of various 
non-autonomous ODEs described in 
Sec. \ref{sec:pullb-attr-tipp} and
\ref{sec:appr-moving-fixed}.
(A) Moving fixed point with $\tilde{x}(t)=-vt$ according to Eqs. (\ref{eq:moving-fixed-point2}) and (\ref{eq:moving-fixed-point-pullback}); $v=1$, $\lambda=1$, $x_{0}=1$ and several initial times $t_{0}$.
(B) Moving fixed point with $\tilde{x}(t)=\mu(-t)^{1/2}$, $t\leq 0$; $\mu=1$, $\lambda=1$, $x_{0}=1$ and several initial times $t_{0}$.
(C) Moving fixed point with changing eigenvalue according to (\ref{eq:jump}),
with $\tilde{x}(t)=\frac{1}{2}r'(t)=v(-t)^{1/2}$, $t\le 0$; $v=1$, $x_{0}=1$, and several initial times $t_{0}$.
}
\label{fig:1}
\end{figure}

The plots in Fig. \ref{fig:1} also illustrate that $\varphi_{pb}(t)$ is a more appropriate characterization of the ``long-time limiting behavior'' than $\varphi_{f}(t_{0})$: namely, for a system with some initial condition, the latter tells us only about its behavior in the infinite future. In order to observe the long-term behavior of the system without going into the infinite future, one has to instead start the system in the infinite past. 
 

If $\lambda$ is very large, the relaxation toward $\tilde{x}(t)$ occurs on a much faster timescale than the dynamics of the moving fixed point $\tilde{x}(t)$ itself. In this case, Eq.\ \ref{eq:moving-fixed-point-pullback} becomes $\varphi_{pb}(t;\lambda)\simeq\tilde{x}(t)$. 

Figs. \ref{fig:1}A and \ref{fig:1}B illustrate two examples of $\tilde{x}(t)$: a linear case, $\tilde{x}(t)=-vt$, and a nonlinear case: $\tilde{x}(t)=\mu(-t)^{1/2}$, where $t\leq 0$. The corresponding solutions are 
\begin{align}
  \varphi^{\text{lin}}(t\mid x_{0},t_{0}) &= (x_{0}+vt)e^{-\lambda (t-t_{0})}+\tilde{x}(t)+\frac{v}{\lambda}\left[1-e^{-\lambda(t-t_{0})}\right] \label{eq:moving-fixed-ex1}, \\
  \varphi^{\text{nonl}}(t\mid x_{0},t_{0}) &= \left(x_{0}-\mu\sqrt{-t_{0}}\right)e^{-\lambda(t-t_{0})}+\tilde{x}(t)+\mu e^{-\lambda t}\int_{|t|}^{|t_{0}|}e^{-\lambda\tau}\dif{\sqrt{\tau}}.   \label{eq:moving-fixed-ex2}
\end{align}
In both cases, $x(t)\simeq \tilde{x}(t)+C$ for large $(t-t_{0})$, where $C$ is a positive constant. For the linear case, $C=v\lambda^{-1}$; for the nonlinear case, $C\simeq (\mu/2)\sqrt{\pi/\lambda}$, the last term in Eq.\ \ref{eq:moving-fixed-ex2}. We conclude that for a system with a moving fixed point, the long-term dynamics follows the fixed point $\tilde{x}(t)$ plus a constant, represented by the last term in Eq.\ \ref{eq:moving-fixed-point2}, which is the displacement of $\tilde{x}(t)$ within the characteristic time $\lambda^{-1}$.

	To briefly summarize, there are three key features of the pullback attractor: if it exists, then it is unique, and it is a close counterpart of the corresponding autonomous ODE attractor under a small perturbation.  It is also important to keep in mind, that a pullback attractor needs not be a single curve of $t$. It can be an envelope.  One such example is $\tfrac{\rd x}{\rd t} = x-t^2x^2$, the pullback attractor of which in $x_0\ge0$ is the set $[0,\frac{1}{(t-1)^2+1}]$ \cite{Robinson2012attractors}.

\subsection{Moving fixed point that vanishes with vanishing
eigenvalue}
\label{sec:moving-fixed-point}

Now let us consider a situation in which the eigenvalue at the moving fixed point is also changing with time: $\deriv{x}{t}=-r'(t)(x-\tilde{x}(t))$ with $r(0)=0$. The solution to this is
\begin{align}
  \label{eq:jump}
  x(t) &= \left(x(0)-\tilde{x}(0)\right)e^{-r(t)}+\tilde{x}(t)-e^{-r(t)}\int_{0}^{t}e^{r(\tau)}\dif{\tilde{x}(\tau)}.
\end{align}
The first term disappears if we let $x(0)=\tilde{x}(0)$. Figure \ref{fig:1}C shows an example in which we take $\tilde{x}(t)=\frac{1}{2}r'(t)=v(-t)^{1/2}$. In this case,
\begin{align}
  \label{eq:jump2}
  x(t) &= (v\hat{t})e^{-r(t)}+2(v\hat{t})^{2}e^{-r(t)}\int_{0}^{t/\hat{t}}e^{r(s\hat{t})}(1-s)\dif{s}.
\end{align}
and $r(t)=\frac{4v}{3}(-t)^{3/2}$ becomes imaginary when $t>\hat{t}$. The fixed point vanishes at $t=0$ at the same
time when the eigenvalue becomes zero.   This is the 
breaking time in the dynamic saddle-node bifurcation (see
below).

\subsection{System with a ``hidden catastrophe''}
\label{sec:pullb-attr-syst}

\begin{figure}
\begin{tabular}{cc}
\includegraphics[width=0.5\linewidth]{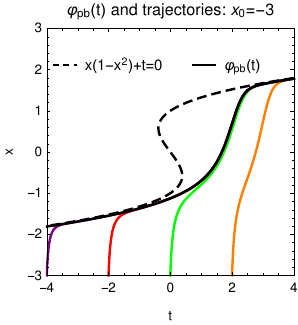}&
\includegraphics[width=0.5\linewidth]{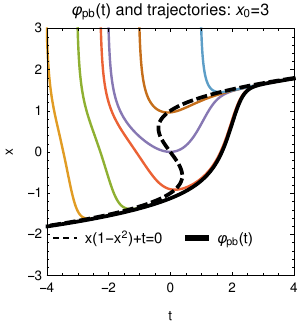} \\
{\bf \Large (A)} & {\bf \Large (B)}
\end{tabular}
\caption{ 
The pullback attractor $\varphi_{pb}(t)$ of $\deriv{x}{t}=x(1-x^{2})+t$ as the bold solid line, along with various trajectories. The dashed line $x(1-x^2)+t=0$ is the bifurcation diagram of the underlying autonomous system when $t$ is treated as bifurcation parameter.
Note how $\varphi_{pb}(t)$ does not jump at $t=0$, but rather changes smoothly to connect the two roots for $t\rightarrow \pm \infty$.
(A) Trajectories with $x_{0}=-3$ for several different $t_{0}$.  
(B) Trajectories with $x_{0}=3$ for several different $t_{0}$.}
\label{fig:2}
\end{figure}

What happens if the vector field $g(x,t)$ of a non-autonomous system $\deriv{x}{t}=g(x,t)$ has multiple roots for a certain range of $t$? The pullback attractor in this case resembles the bifurcation diagram of an autonomous system $\deriv{x}{t}=g(x,\lambda)$ with a fold catastrophe: namely, it connects the root of $g(x,t)$ when $t\to -\infty$ with the root of $g(x,t)$ when $t\to +\infty$. Fig. \ref{fig:2} shows the pullback attractor of an example system 
\begin{equation}
     \deriv{x}{t}=x(1-x^{2})+t,
\label{eq15}
\end{equation}
along with several sample trajectories for several initial conditions. The curve $x(1-x^{2})+t=0$ is also plotted to illustrate how the pullback attractor tracks the autonomous bifurcation diagram for $t\to\pm\infty$. Note trajectories do not immediately jump to the other branch at $t=0$, and this is the non-autonomous counterpart of the bottleneck feature.

\section{Dynamic saddle-node bifurcation}
\label{sec:dynamic-saddle-node}

Tipping point phenomena are particularly well-represented by one-dimensional dynamical systems with saddle-node bifurcations due to their structural stability and robustness.  In the context of 
non-autonomous dynamics, the problems deal with a moving fixed point which can also change its character (\emph{e.g.}, eigenvalue), as well as disappear altogether.  
In Sec.\ \ref{sec:non-auton-syst}, we established simple descriptions of three different mathematical phenomena: (\emph{i}) relaxation of trajectories towards a moving stable fixed point; (\emph{ii}) relaxation of trajectories towards a moving stable fixed point with vanishing eigenvalue; (\emph{iii}) the effect of a saddle-node ghost in non-autonomous dynamical systems. We shall now focus on a simple non-autonomous equation that exhibits these three types of behavior, as well as the time-dependent equivalent of a saddle-node bifurcation---we shall say that such a system exhibits a \emph{dynamic saddle-node bifurcation}.

\subsection{Model for a dynamic saddle-node bifurcation}
\label{sec:non-auton-saddle}

\begin{figure}
\begin{tabular}{cc}
\includegraphics[width=0.49\linewidth]{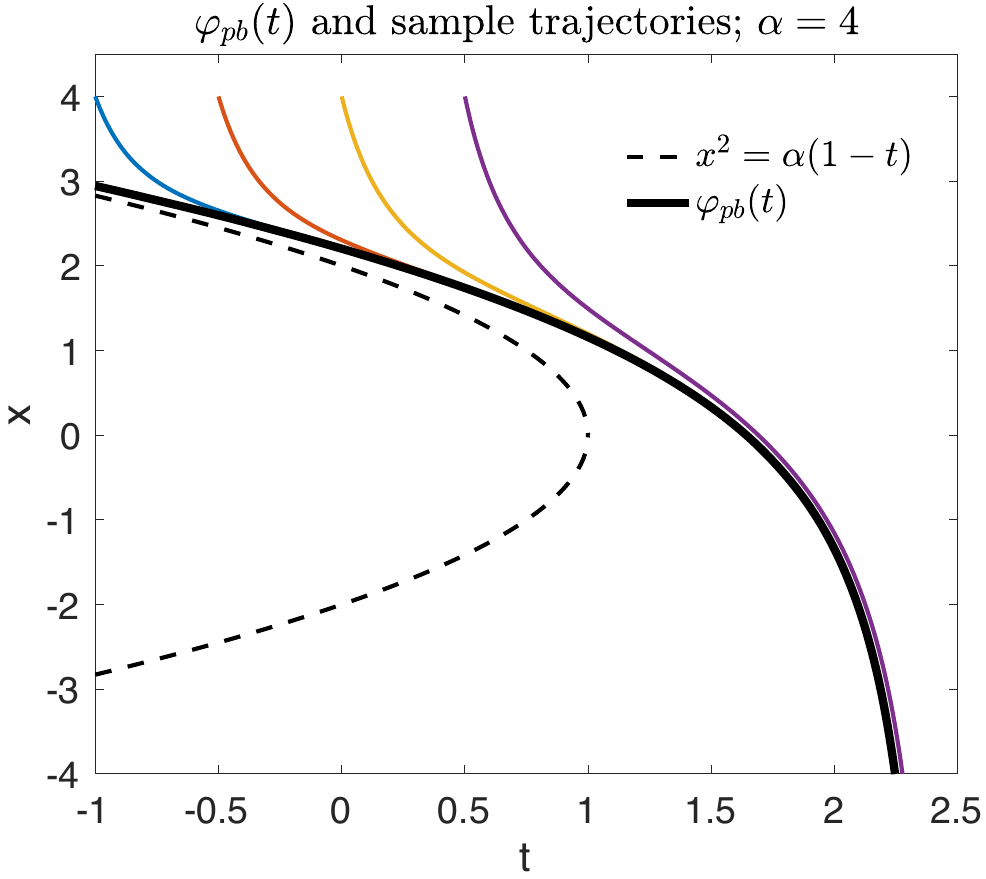} &
\includegraphics[width=0.5\linewidth]{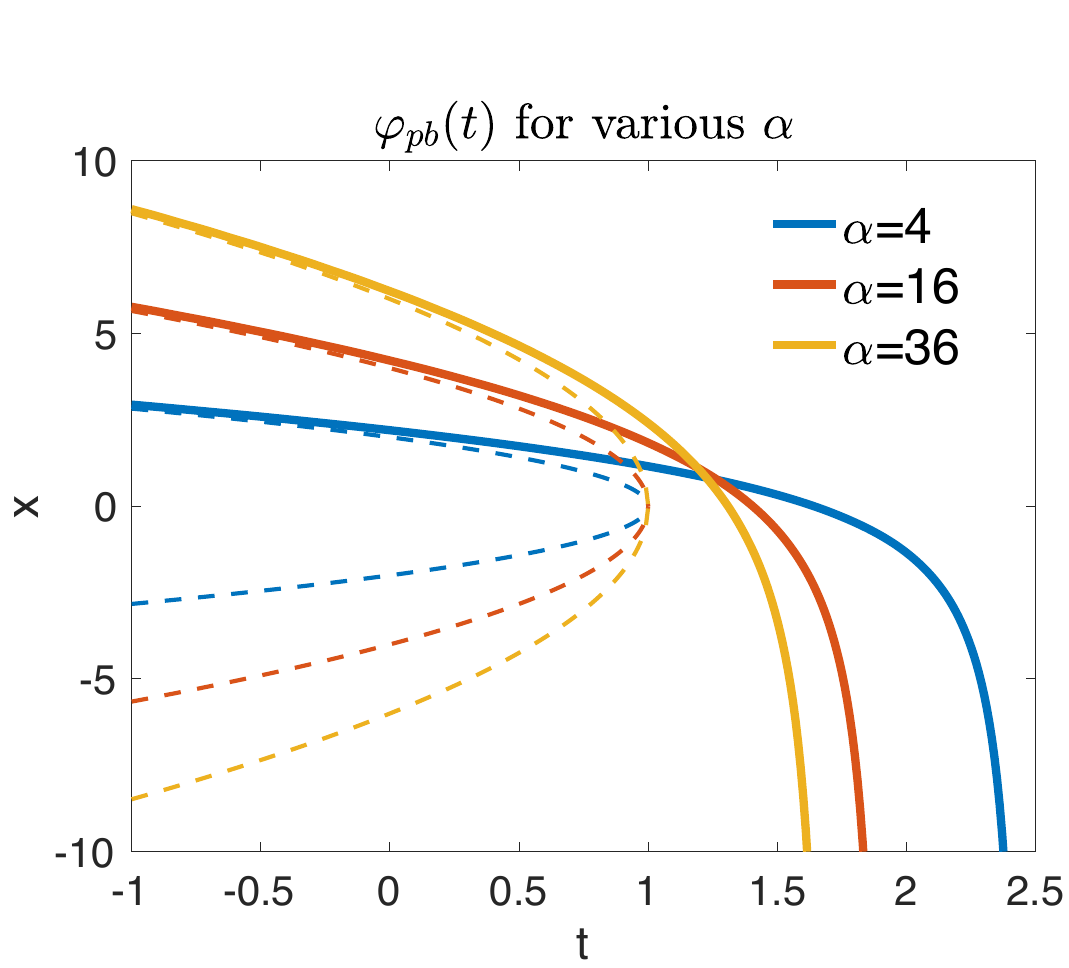}\\
{\bf \Large (A)} & {\bf \Large (B)}
\end{tabular}
\caption{(A) Plot of $\varphi_{pb}(t)$ and trajectories with $\alpha=4$, $x_{0}=4$ for several different $t_{0}$. The dashed line $x^2=\alpha(1-t)$ is the bifurcation diagram of the underlying autonomous system when $t$ is treated as bifurcation parameter.
(B) Plot of the pullback attractor $\varphi_{pb}(t)$ for different values of $\alpha$. Note how the point of no return $t^{*}$ decreases with $\alpha$ increases. Thin dashed lines are again
the corresponding ``autonomous bifurcation diagrams'' when $t$ is treated as bifurcation parameter.
}
\label{fig:3}
\end{figure}

A simple example in biology is given by the relationship between many animal population densities and the per capita growth rate of the population---specifically, the \emph{strong Allee effect} describes how the per capita growth rate of some populations is maximal at some intermediate density and negative at low densities. Thus, for a laboratory experiment involving, for example, yeast, one could map the bifurcation diagram of per capita growth rate (the state variable) vs.\ a changing population density (the parameter) by performing regular dilutions on the populations (as done experimentally by Dai \emph{et.\ al} \cite{jgore1,jgore2}). Then as the dilution factor is increased, the system passes its \emph{breaking point} at the exact time when the population density decreases beyond the critical point where the instantaneous per capita growth rate is no longer sufficient to maintain the population, and the system begins to crash. Since the regular dilution and dynamics happens at the same time, it is necessary to consider the time-dependent parameter in the saddle node bifurcation. 

We now introduce a non-autonomous dynamical system with a dynamic saddle-node bifurcation. Consider the differential equation for a generic saddle-node bifurcation dynamical system in the case when the bifurcation parameter $p$ is an explicit linear function of time: $p(\tau)=p_{0}-V\tau$ where $p_{0},V > 0$, so that we have 
\begin{align}
  \label{eq:model-untrans}
  \deriv{X}{\tau} &= -\lambda X^{2}+(p_{0}-V\tau),
\end{align}
where $V$ is the ``speed'' with which the parameter is changing. Here $V$ is not necessarily small. This generic equation can be considered as the local approximation around the bifurcation point for this biology example within the time range of interest.

This type of dynamic saddle-node bifurcation in fact is very common. Another motivating model  introduced by Held and Suarez, and Fraedrich 
\cite{held-suarez,fraedrich} describes the evolution of the global mean surface temperature of an ocean-covered spherical planet subjected to radiative heating; it also takes into account ice-albedo and greenhouse feedback. This model, where $T$, the mean temperature, is the state variable, is given by \cite{kktung}
\begin{align}
  \label{eq:fraedrich-orig}
  c\deriv{T}{\tau} &= R_- - R_+,
\end{align}
where $R_-$ and $R_+$ represent the incoming and outgoing radiation respectively and are given in terms of physical parameters:
\begin{align}
  \label{eq:radiation}
  R_- &= \frac{1}{4}\mu I_{0}(1-\alpha_{p}) \qquad R_+ = \epsilon_{sa}\sigma T^{4},
\end{align}
where $I_{0}$ is the solar constant, $c>0$ is the thermal inertia of the ocean, $\alpha_{p}$ is the planetary albedo, $\epsilon_{sa}$ is the effective emissivity, and $\mu$ is an external parameter which allows for variations in the solar constant or for long-term variations in planetary orbit. Now, consider a planet whose albedo-temperature relation's slope is negligible, and whose albedo increases (due to some unrelated cause) at a constant rate $\gamma$ in time. Then the planetary albedo is given by 
\begin{align}
  \label{eq:albedo}
  \alpha_{p}(t) &= \gamma \tau
\end{align}
Then, Eq.\ \ref{eq:fraedrich-orig} can be written as 
\begin{align}
  \label{eq:trans-fraedrich}
  \deriv{T}{\tau} &= -\lambda T^{4}+p_0-p_0\gamma \tau,
\end{align}
where 
\begin{align}
  \label{eq:trans-rules}
  \lambda \equiv \frac{\epsilon_{sa}\sigma}{c} \qquad & p_0\equiv \frac{I_{0}\mu}{4c}.
\end{align}
Note that $\lambda>0$. Then by introducing a transformed variable $X\equiv T^{2}$, we have 
\begin{align}
  \label{eq:doubly-trans}
  \deriv{X}{\tau} &= 2T\deriv{T}{\tau} = 2(-\lambda X^{2}+p_0-p_0\gamma \tau)\sqrt{X}.
\end{align}
Presuming that we are bounded away from domains that might cause existence problems, Eq.\ \ref{eq:doubly-trans} is bifurcation-equivalent to 
\begin{align}
  \label{eq:reduced-fraedrich}
  \deriv{X}{\tau} &= -\lambda X^{2}+p_0-p_0\gamma \tau,
\end{align}
which is in the form of our model, Eq.\ \ref{eq:model-untrans}. In the context of climate science terminology, then, our model is a simple example of bifurcation-induced tipping (B-tipping) where the parameter change is not necessarily slow.

 One can introduce transformed variables $x$ and $t$ alongside the corresponding parameter $\alpha$:
\begin{equation}
  \label{eq:change-var}
  x = \frac{\lambda p_{0}X}{V}, \quad t=\frac{V\tau}{p_{0}}, \quad \text{and} \quad \alpha = \lambda p_{0}^{3}V^{-2}.
\end{equation}
Then we arrive at a ``non-dimensionalized'' version of Eq.\ \ref{eq:model-untrans}
\begin{align}
  \label{eq:model-trans}
  \deriv{x(t)}{t} &= -x^{2}+\alpha(1-t). 
\end{align}
Denoting the right-hand-side of Eq.\ \ref{eq:model-trans} as $f(x,t)\equiv -x^{2}+\alpha(1-t)$, it is easy to see that for any $t<1$, the vector field $f(x,t)$ has two roots---at $x_+(t)=\sqrt{\alpha(1-t)}$ and $x_-(t)=-\sqrt{\alpha(1-t)}$. These two roots approach each other 
as $t$ increases, and annihilate at $\hat{t}=1$, and for all $t>1$, $f(x,t)$ is always negative: all trajectories eventually go to $-\infty$. 

Fig. \ref{fig:3} shows several solutions to Eq.\ \ref{eq:model-trans} as well as the bifurcation diagram for the corresponding autonomous system assume $t$ as the bifurcation parameter.  Different initial conditions converge to a single ``master
curve'' | this phenomenon is  {\em backward attraction}
\cite{kloeden-rasmussen}.
Here we pause to note that the trajectories of these dynamics describe systems with tipping points---they follow a stable equilibrium until a sudden and abrupt transition into a state very different from the original because of some change in its surrounding environment; this is caused by the sudden disappearance of the stable fixed point in $f(x,t)$ at $t=\hat{t}$.  We shall call $\hat{t}$ the {\em breaking time}.

However, it is important to note that  while its environment has
already fundamentally changed, a dynamical system does not have an immediate drastic response at exactly the time $\hat{t}$, when $f(x,t)$ goes from having two fixed points (for $t<\hat{t}$) to none (for $t>\hat{t}$). Rather, there is a delay as the system traverses the bottleneck caused by the saddle-node ghost present in $f(x,t)$, until it finally reaches $-\infty$ at a time $t^{*}>\hat{t}$ (see Fig. \ref{fig:3}A): we shall call this time the ``point of no return''---this choice of naming will become clearer shortly. 
 Since the transition occurs later than the static saddle node bifurcation point $\hat{t}$, it is also called a {\it delayed bifurcation} \cite{Haberman1979, Rachel2015}. 

Fig.\ \ref{fig:3} clearly shows that the time interval between the breaking time and point of no return (\emph{i.e.,} $(\hat{t},t^{*})$ ) for a system can be
quantified in terms of the system's pullback attractor. Indeed, since tipping point problems are concerned with systems which have in the past followed a stable equilibrium, their dynamics are suitably captured by the pullback attractor of the system. 

An analytical solution can be found for Eq.\ \ref{eq:model-trans}, as we show in Sec.\ \ref{sec:airy-sol}.
What makes a non-autonomous treatment of such systems useful compared to the standard, autonomous models, is that it allows an 
assessment of the point of no return for a system with parameters that change on a similar time scale to the phase space dynamics. On the other hand, even for the slow-varying systems,  an autonomous treatment  is not valid any more, because the point of no return $t^{*}$ is sufficiently big such that the parameter of the system has changed significantly. 

Note that trajectories that followed the stable equilibrium in the past behave exactly as the pullback attractor, and experience the same difference between $\hat{t}$ and $t^{*}$---it is precisely this fact that makes the pullback attractor so vital in the study of tipping point phenomena, since systems which have not yet tipped are assumed to have been following some equilibrium in the past. Fig. \ref{fig:3}B illustrates the dependence of $t^{*}$ as determined by $\varphi_{pb}(t;\alpha)$ on $\alpha$; or, in the untransformed variables, the dependence on the velocity $V$ and starting value $p_{0}$ of the parameter. In addition, it is always true that the point of no return $t^{*}$ is larger than the breaking time $\hat{t}$.  Thus we see that the critical transition is delayed due to the system being bottlenecked, and that this bottlenecking depends on $\alpha$.

\begin{figure}[h]
\[
\includegraphics[width=2.6in]{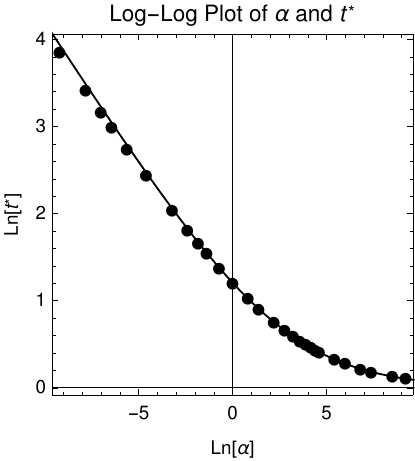}
\]
\caption{The time of the point of no return, $t^*$, 
as a function of $\alpha$,
in a log-log plot, according to Eq.\ \ref{eq:model-trans}.
The points are obtained from numerical solutions to
the differential equation, e.g., from Fig. \ref{fig:3}B.  
There are two distinct regimes with 
$\alpha<1$ and $\alpha>1$.  The thin solid curve $t^*=1+2.338\alpha^{-\frac{1}{3}}$ is obtained from
the analytical solution of the pullback attractor.
For a very slowly changing $p(t)$,
$\alpha\gg 1$ and $t^*\simeq\hat{t}=1$.  
}
\label{fig:4}
\end{figure}

Fig.\ \ref{fig:4} shows a log-log plot for the point of no return $t^*$ as the function of $\alpha$. Note that the speed with which the parameter is changing ($V$), and the initial ``distance'' from the bifurcation ($p_{0}$), contribute to $\alpha$ differently: $\alpha\sim p_{0}^{3}V^{-2}$. Furthermore, note the difference in behavior for the regimes $\alpha>1$ and $\alpha<1$. In particular, For a very slowly changing $p(t)$, $\alpha\gg 1$ and $t^*\simeq\hat{t}=1$.

\subsection{Analytical solution in terms of Airy functions}
\label{sec:airy-sol}
Eq.\ \ref{eq:model-trans},
\begin{equation}
 \frac{\rd x(z)}{\rd z} = -x^2+\alpha (1-z),
\label{riccati}
\end{equation}
is a Riccati equation, $\tfrac{\rd x}{\rd t} =
a_0(t)+a_1(t)x+a_2(t)x^2$, with 
$a_2(t)=-1$, $a_1(t)=0$, and $a_0=\alpha (1-t)$.  
According to the standard method of attack, this
1\textsuperscript{st} order nonlinear ODE can be transformed into 
a 2\textsuperscript{nd} order linear ODE via 
$x=\tfrac{1}{u}\left(\tfrac{\rd u}{\rd t}\right)$ so that we have
\begin{equation}
\label{eq:airyode}
     \frac{\rd^2 u}{\rd t^2} =
        \alpha (1-t) u.
\end{equation}
Eq.\ \ref{eq:airyode} is related to the Airy equation, and can be solved in terms of Airy functions. We are 
interested in its solution with initial values $u(t_0)=1$ and
$\tfrac{\rd u(t_0)}{\rd t}=x_0$.

\begin{figure}[h]
\[
\includegraphics[width=3.in]{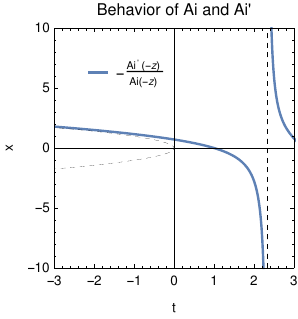}
\]
\caption{The left dashed curve is
$-\tfrac{1}{\text{Ai}(-z)}\tfrac{\rd\text{Ai}(-z)}{\rd z}$
as function of $z$.
The solid curve crosses the horizontal axis 
at $z=1.02$ since $\text{Ai}(-1.02)$ is a local 
maximum of the Airy function.  It has an asymptote
(vertical dashed line) at $z=2.34$ since
$\text{Ai}(-2.34)=0$.  The dashed parabola represents
 $\pm\sqrt{-z}$.  Note that the solid line asymptotically
approaches to the vertical dashed line since 
$\text{Ai}(z) \sim \tfrac{1}{2}\pi^{-\frac{1}{2}}
z^{-\frac{1}{4}}e^{-\frac{2}{3}z^{3/2} }$ as
$z\rightarrow\infty$ \cite{bender-orszag}.  Thus, the 
solid line goes as $\sqrt{-z}-(4z)^{-1}$ when $z\rightarrow-\infty$.
}
\label{pb-attractor}
\end{figure}

The general solution to Eq.\ \ref{eq:airyode} is
\begin{equation}
    u(t) = \Big[ C_1 \text{Ai}(z)
            + C_2 \text{Bi}(z) \Big]_{z=\sqrt[3]{\alpha}(1-t)} ,
\end{equation}
where $C_{1,2}$ are two constants, and 
\begin{subequations}
\begin{equation}
  \text{Ai}(z) = \frac{1}{\pi}\int_0^{\infty}\cos\left(
                  \frac{\xi^3}{3}+z\xi\right)\rd\xi, 
\end{equation}
\begin{equation}
     \text{Bi}(z) = \frac{1}{\pi}\int_0^{\infty}\left[
              \exp\left(-\frac{\xi^3}{3}+z\xi\right) + \sin\left(
                  \frac{\xi^3}{3}+z\xi\right) \right] \rd\xi. 
\end{equation}
\end{subequations}
Then,
\begin{equation}
   \frac{\rd u(t) }{\rd t} 
         =  -\sqrt[3]{\alpha} \left[ C_1 
       \tfrac{\rd\text{Ai}(z)}{\rd z} 
            + C_2 \tfrac{\rd\text{Bi}(z)}{\rd z} \right]_{z=\sqrt[3]{\alpha}
                 (1-t)},
\end{equation}
and the solution to Eq.\ \ref{riccati} is
\begin{equation}
    x(t) =  -\sqrt[3]{\alpha}
         \left[ \frac{\displaystyle
               C_1\frac{\rd\text{Ai}(z)}{\rd z}
            + C_2 \frac{\rd\text{Bi}(z)}{\rd z}  }{C_1\text{Ai}(z)
            + C_2 \text{Bi}(z) }
            \right]_{ z=\sqrt[3]{\alpha}(1-t)},
\end{equation}
with the constants
\begin{subequations}
\begin{equation}
  C_1(t_0) = \left[ \frac{ \tfrac{\rd\text{Bi}(z)}{\rd z}
              +\frac{x_0}{\sqrt[3]{\alpha}} \text{Bi}(z)}{ \text{Ai}(z)\tfrac{\rd\text{Bi}(z)}{\rd z}
                 -\text{Bi}(z)\tfrac{\rd\text{Ai}(z)}{\rd z}   }
             \right]_{z=\sqrt[3]{\alpha}(1-t_0)},
\end{equation}
\begin{equation}
  C_2(t_0) = \left[ \frac{ -\frac{x_0}{\sqrt[3]{\alpha}} \text{Ai}(z)
               -\tfrac{\rd\text{Ai}(z)}{\rd z}
              }{ \text{Ai}(z)\tfrac{\rd\text{Bi}(z)}{\rd z}
                 -\text{Bi}(z)\tfrac{\rd\text{Ai}(z)}{\rd z}   }
             \right]_{z=\sqrt[3]{\alpha}(1-t_0)}.
\end{equation}
\end{subequations}
The pullback attractor is obtained when 
$t_0\rightarrow -\infty$. Since $\text{Ai}(z)=0$
and $\text{Bi}(z)\rightarrow +\infty$ as 
$z\rightarrow +\infty$, we have
\begin{equation}
   \lim_{t_0\rightarrow-\infty}
         \left(\frac{C_2(t_0)}{C_1(t_0)}\right)
       = \lim_{z\rightarrow +\infty}
     \left( \frac{ -\frac{x_0}{\sqrt[3]{\alpha}} \text{Ai}(z)
               -\tfrac{\rd\text{Ai}(z)}{\rd z} }{ \tfrac{\rd\text{Bi}(z)}{\rd z}
              +\frac{x_0}{\sqrt[3]{\alpha}} \text{Bi}(z)}
             \right) = 0.
\end{equation}
\begin{equation}
    \varphi_{pb}(t) =  -\left[ \frac{\sqrt[3]{\alpha} }{\text{Ai}(z)}
          \frac{\rd\text{Ai}(z)}{\rd z}
              \right]_{z=\sqrt[3]{\alpha}(1-t)}.
\end{equation}
The behavior of $-\tfrac{1}{\text{Ai}(-z)}\tfrac{\rd\text{Ai}(-z)}{\rd z}$ is shown in Fig. \ref{pb-attractor}.
The denominator $\text{Ai}(z)$ vanishes when $z\approx -2.338$, corresponding to the asymptote where $\varphi_{pb}\rightarrow -\infty$.  Hence the time of no return is given by $t^*\approx 1+2.338\alpha^{-\frac{1}{3}}$: in other words, once the system breaks at $\hat{t}=1$, the amount of time it takes for the system to reach the point of no return and fully tip (\emph{i.e.},  $t^{*}-\hat{t}$) scales according to $\alpha$ (see Fig.\ \ref{fig:4}) is $2.338\alpha^{-\frac{1}{3}}$. In particular, for a very slowly changing $p(t)$, $\alpha\gg 1$ and $t^*\simeq\hat{t}=1$.
For the original system in Eq. \ref{eq:model-untrans}, the time window scales as $\tau^{*}-\hat{\tau}\simeq 2.338(\lambda V)^{-1/3}$.

Similar scaling law can also be obtained through asymptotic approximation of the slowly varying solution and has been verified experimentally in various physical systems. See \cite{Erneux1989, Strogatz1995,Haberman1979,Rachel2015} and references therein.

\section{Discussion}
\label{sec:discuss}




The modern theory of non-autonomous dynamical systems has made  substantial progress in study of the asymptotic behavior subject to time-dependent parameter. However, it has a mathematical setup that is rather different from the theory of autonomous dynamical systems. Most of the existing  literature is fairly technical and almost impenetrable
for all except the specialists. Moreover, the applications of these  modern theories to real world applications are not trivial. The goal of the present paper is to initiate an applied mathematical study of non-autonomous dynamical systems and emphasis on the application of the theory  to problems arising in the applied sciences. In particular, we focus on the theory of pullback attractors and applications of the dynamical saddle-node bifurcations. The future extension of this theory will be to study the parameter experiencing random force and it will lead to the theory of random dynamical system. The counterpart of the pullback attractors will be the random attractors.

With the help of the pullback attractors, we are able to identify the breaking time $\hat{\tau}$ and the point of no return $\tau^*$ in the dynamical saddle-node bifurcations, which is the counterpart of saddle node ghost effect in the autonomous dynamical system.  
The importance of the relation $\tau^{*}>\hat{\tau}$ is that for a changing environment which has just undergone a fundamental change, there remains an interval of time $(\hat{\tau},\tau^{*})$ during which the system can be saved from  asymptotically reaching the different state if the changes to the environment are rapidly reversed. In other words, as long as the point of no return has not yet been reached, the system is still in theory salvageable. This has potential significance in the management of tipping point phenomena, since it provides a way to quantify and estimate the window of time after a significant environment change during which it is still possible to prevent the system from fully tipping. In particular, the smaller the rate of change $V$ or the curvature of bifurcation $\lambda$ is, the bigger the window of time gets. 


In the biology example described in Sec. \ref{sec:non-auton-saddle}, the system is still \emph{salvageable} even it begins to crash after regular dilution, as the point of no return (the time at which the entire population has died off) has not yet been reached. Indeed, if the environment change is rapidly reversed by adding a sufficient number of additional yeast cells into the system, then the population can be saved, and the per capita growth rate will recover. It is easy to imagine using similar models to track and estimate this time frame of salvageability for phenomena such as collapsing fish stocks due to overfishing.

In the climate science example,  the breaking time and the point of no return for the system can be calculated as well. Physically, the breaking time corresponds to the point at which the planet's ocean's environment can no longer support a stable, ``interglacial'' state, and the point of no return corresponds to the point at which the ocean reaches an irreversible ``deep freeze'', e.g., a snowball planet \cite{kktung}. However, if the albedo is quickly decreased during this time window, it is possible that the ``deep freeze" could be saved.

\section*{Acknowledgments}
The authors are indebted to the anonymous referees for the valuable
comments that allowed them to improve a previous version of the
manuscript.
The authors wish to thank Professors Luonan Chen and Rachel Kuske for helpful discussions. The work is partially supported by NIH grant R01-GM109964.

\bibliographystyle{unsrt}
\bibliography{bibliography}

\end{document}